
\magnification=\magstep1
\baselineskip=12pt
\overfullrule=0pt
\font\twelvebf=cmbx12
\def\P#1{P\uppercase\expandafter{\romannumeral #1}}

\rightline{IASSNS-HEP-93/9}
\rightline{Feb 1993}
\vskip .3in
\centerline{\twelvebf B\"acklund Transformations}
\centerline{\twelvebf of MKdV and Painlev\'e Equations}
\vskip .8in
\centerline{\bf Jeremy Schiff}
\smallskip
\centerline{\it Institute For Advanced Study}
\centerline{\it Olden Lane, Princeton, NJ 08540}
\vskip .8in
\centerline{\bf Abstract}
\smallskip
{\narrower\smallskip
\noindent For $N\ge 3$ there are $S_N$ and $D_N$ actions on the
space of solutions of the first nontrivial equation in the $SL(N)$ MKdV
hierarchy, generalizing the two $Z_2$ actions on the space of solutions
of the standard MKdV equation. These actions survive scaling
reduction, and give rise to transformation groups for certain (systems of)
ODEs, including the second, fourth and fifth Painlev\'e equations.\smallskip}
\vskip .8in
\noindent Given a solution $j$ of the MKdV equation
$$ j_t=j_{xxx}-{\textstyle{3\over 2}}j^2j_x  \eqno{(1)}$$
we can construct new solutions, $-j$ and $j-{2\over q}$, where $q$
satisfies
$$ \eqalign{q_x+qj&=1\cr
            q_t+q(j_{xx}-{\textstyle{1\over 2}}j^3)&=
               (j_x-{\textstyle{1\over 2}}j^2). } \eqno{(2)}$$
Equations (2) constitute a strong auto-B\"acklund transformation for the
MKdV equation, distinct from the usual one given in the literature (see
for example [1], Chapter 8, Exercise 2), and discovered, I believe,
in the context of Painlev\'e analysis [2]. If we choose the integration
constant arising in the solution of (2) appropriately,
the square of this transformation
is the identity; but when combined with the
$j\rightarrow -j$ transformation it can, generically, be used to
generate an infinite number of solutions from a particular one (solutions
of MKdV that are periodic under the action of the combined symmetry are
discussed in [3]).

Unlike the standard auto-B\"acklund transformation for
MKdV, the transformations $j\rightarrow -j$, $j\rightarrow j-{2 \over q}$
do not contain dimensionful parameters,
and hence survive scaling reduction.\footnote{$^1$}{When the dimensionful
parameter in the standard transformation is set to zero, the transformation
becomes trivial. The standard auto-B\"acklund transformation for KdV [1]
remains non-trivial even when the dimensionful parameter is set to zero,
and this was used in [4], along with the Miura map, to obtain a
different derivation of the transformation group of \P2 from
the one we are about to see.}
Setting $j=-2(3t)^{-{1\over 3}}J(w)$ where
$w=-x(3t)^{-{1\over 3}}$, (1) reduces to the second Painlev\'e equation (\P2)
$$ J''=2J^3+Jw+\alpha \eqno{(3)}$$
where a prime denotes differentiation with respect to $w$ and $\alpha$ is an
integration constant. Setting $q=(3t)^{1\over 3}Q(w)$ in (2), we can solve for
$Q$ and thus we have two {\it explicit} transformations for (3),
$$ \eqalign{&J\rightarrow -J,~~~~~~~~\alpha\rightarrow -\alpha \cr
            &J\rightarrow J+{{{\textstyle{1\over 2}}-\alpha}\over
                             {J'-J^2-{\textstyle{1\over 2}}w}},~~~~~~~~
                       \alpha \rightarrow 1-\alpha. } \eqno{(4)}$$
These transformations, which both square to the identity, generate the
well-known transformation group of \P2 (see [5],[6] and references therein).
In solving for $Q$ I have assumed $\alpha\not={1\over 2}$; for $\alpha=
{1\over 2}$ there is a one parameter family of solutions of \P2 given by
the solutions of $J'=J^2+{1\over 2}w$ (which can be solved in terms of
Airy functions).

The purpose of this note is to present generalizations of the above
transformations for the SL(N) MKdV equations, $N\ge 3$. The above
transformations for the standard MKdV equation actually extend to the
MKdV hierarchy, and similarly in the $SL(N)$ case the relevant transformations
extend to the hierarchy. But for clarity we will focus just on the lowest
nontrivial equation in the hierarchy.
On scaling reduction all the transformations become
explicit. The (lowest nontrivial) $SL(3)$ and $SL(4)$ MKdV equations reduce,
respectively, to the \P4 and \P5 equations (the first of these facts
originated, I believe, in [7]; the second is, I believe,
new); we recover the transformation groups investigated by Okamoto
for \P4 [6] (see also [8]) and \P5 [9].

The $SL(N)$ MKdV hierarchy describes evolutions of $N$ fields $j_i$,
$i=1,..,N$, with $\sum_{i=1}^{N}j_i=0$. Writing $\Sigma={1\over N}
\sum_{i=1}^{N}j_i^2$, the lowest nontrivial evolution in the hierarchy is
$$ \partial_tj_i=
      \partial_x\left[ \sum_{r=1}^{N-1}\left(1-{\textstyle{{2r}\over N}}\right)
                                 \partial_x j_{i+r({\rm mod}~N)}
               + j_i^2 - \Sigma \right],
        ~~~~~~i=1,..,N\eqno{(5)}$$
or, equivalently,
$$ \partial_t(j_i-j_{i+1})=
  \partial_x[\partial_x(j_i+j_{i+1})+j_i^2-j_{i+1}^2],~~~~~~~~i=1,..,N-1.
  \eqno{(5)^\prime}$$
A simple way to obtain (5) is by reduction of the $SL(N)$ self-dual Yang-Mills
equations with an ansatz given in the appendix. (5) has one obvious symmetry
group:

{\narrower\smallskip
\noindent{\it Prop.1: $D_N$ Invariance of (5).}

\noindent Equations (5) are invariant under the $D_N$ action generated by
$$\eqalign{&A:j_i\rightarrow j_{i+1({\rm mod}~N)},x\rightarrow x,
                   t\rightarrow t \cr
           &B:j_i\rightarrow j_{N+1-i},x\rightarrow-x,
                   t\rightarrow-t} \eqno{(6)}$$
which satisfy $A^N=B^2=I$, $ABAB=I$.\smallskip}

\noindent The other symmetry group is less obvious:

{\narrower\smallskip
\noindent{\it Prop.2: $S_N$ Invariance of (5).}

\noindent There is an $S_N$ action on solutions
of (5); the action of the fundamental transpositions $T_i=(i~i+1)$,
$i=1,..,N-1$, is given by
$$\eqalign{j_i&\rightarrow j_i+q_i^{-1} \cr
           j_{i+1}&\rightarrow j_{i+1}-q_i^{-1} \cr
           j_r&\rightarrow j_r,~~r\not=i,i+1 }\eqno{(7)}$$
where $q_i$ satisfies\footnote{$^2$}{Equations (8) determine $q_i$ up to a
parameter. The precise action of $T_i$ is determined by picking a
suitable boundary condition satisfied by the solution $j_1,...,j_N$ on which
we wish to act, and requiring the $T_i$ to preserve this condition. For
example, for the scaling reduction of (5) we will shortly consider, we
fix this parameter by requiring that $q_i$ should also have a well-defined
scaling behavior.}
$$ \eqalign{&q_{ix}+(j_i-j_{i+1})q_i +1=0\cr
            &q_{it}+[(j_i+j_{i+1})_x+j_i^2-j_{i+1}^2]q_i
                   +(j_i+j_{i+1}) =0.} \eqno{(8)}$$
\smallskip}

\noindent For a complete understanding of the origin of
these transformations, and
why the group generated by the transformations $T_i$ is $S_N$,
the reader is referred to [10]. The basic argument however is quite simple:
consider the $N$th order differential operator
$L=(\partial+j_N)..(\partial+j_2)(\partial+j_1)$, and choose a basis
$\{\psi_1,..,\psi_N\}$ for the kernel of $L$ such that
$\{\psi_1,..,\psi_i\}$ is a basis
for the kernel of $(\partial+j_i)..(\partial+j_2)(\partial+j_1)$ for
each $i$, $i=1,..,N$. It is easy to check that switching $\psi_i$ and
$\psi_{i+1}$ induces the change in the $j_r$ given by (7), with $q_i$
satisfying the first equation in (8); the second equation in (8) is then
deduced directly from (5).

There are  relations between the $D_N$ and $S_N$ generators; two
obvious ones are
$$\eqalign{ AT_i &= T_{i-1}A,~~~~~~~i=2,3,..,N-1\cr
            BT_i &= T_{N-i}B,~~~~~~i=1,2,..,N-1. } \eqno{(9)}$$
It is natural to define a transformation $T_N$ by $T_N\equiv A^{-1}T_{N-1}A$;
this satisfies $AT_1=T_NA$ and  $BT_N=T_NB$.
The explicit action of $T_N$ is
$$\eqalign{j_N&\rightarrow j_N+q_N^{-1} \cr
           j_1&\rightarrow j_1-q_N^{-1} \cr
           j_r&\rightarrow j_r,~~r\not=N,1 }\eqno{(10)}$$
where $q_N$ satisfies
$$ \eqalign{&q_{Nx}+(j_N-j_1)q_N +1=0\cr
            &q_{Nt}+[(j_N+j_1)_x+j_N^2-j_1^2]q_N
                   +(j_N+j_1) =0.} \eqno{(11)}$$
It must be emphasized that $T_N$ is {\it not} a pure $S_N$ transformation,
and should not be confused with the fundamental transposition $(1~N)$ in $S_N$,
which generically changes all the $j_i$. Having introduced $T_N$, it is
clear that the transformation group for (5) is a
semi-direct product of the group generated by $T_1,..,T_N$ with
the group $D_N$.

We now consider the scaling reduction of (5).
Writing $j_i=t^{-{1\over 2}}J_i(w)$ where $w=t^{-{1\over 2}}x$, we
find that we can at once integrate each equation of (5) to obtain  the
reduced system
$$ -{\textstyle{1\over 2}}wJ_i+\alpha_i=
      \sum_{r=1}^{N-1}\left(1-{\textstyle{{2r}\over N}}\right)
      J_{i+r({\rm mod}~N)}^{\prime} +  J_i^2 - S,
        ~~~~~~i=1,..,N.\eqno{(12)}$$
Here  $S={1\over N}\sum_{r=1}^N J_r^2$, a prime denotes differentiation
with respect to $w$,the $\alpha_i$, $i=1,..,N$, are constants satisfying
$\sum_{r=1}^N\alpha_i=0$, and $\sum_{r=1}^NJ_i=0$. Because of the square
roots in the reduction formulae, (12) displays a residual scale invariance
$w\rightarrow
-w$, $J_i\rightarrow -J_i$.\footnote{$^3$}{Similar considerations give rise to
the extra invariances of equation (3) under $J\rightarrow\lambda J$,
$w\rightarrow\lambda^2w$ with $\lambda^3=1$.} This can be eliminated
by setting $J_i(w)=w^{-1}K_i(z)$,
where $z=w^2$, to obtain the system
$$ (\alpha_i-{\textstyle{1\over 2}}K_i)z
   =\sum_{r=1}^{N-1}\left(1-{\textstyle{{2r}\over N}}\right)
      (2z\dot{K}_{i+r({\rm mod}~N)}
     -K_{i+r({\rm mod}~N)}) + K_i^2 - T,~~~~~~i=1,..,N. \eqno{(13)}$$
Here a dot denotes differentiation with respect to $z$,
$T={1\over N}\sum_{r=1}^NK_i^2$, and $\sum_{r=1}^NK_i=0$.
We could, of course, have obtained (13) directly from (5) by substituting
$j_i=x^{-1}K_i(z)$ where $z=t^{-1}x^2$, but if we do this it is somewhat
harder to see the integrations that can be done.

Under scaling reduction (i.e. setting $q_i=xQ_i(z)$) we find we can solve (8)
for $q_i$; we can thus write down both the $D_N$ and $S_N$ actions explicitly:

{\narrower\smallskip
\noindent{\it Prop.1$^\prime$: $D_N$ Invariance of (13).}

\noindent Equations (13) are invariant under the $D_N$ action generated by
$$\eqalign{&A:K_i\rightarrow K_{i+1({\rm mod}~N)},z\rightarrow z,
                   \alpha_i\rightarrow \alpha_{i+1({\rm mod}~N)} \cr
           &B:K_i\rightarrow -K_{N+1-i},z\rightarrow-z,
                   \alpha_i\rightarrow-\alpha_{N+1-i}} \eqno{(14)}$$
which satisfy $A^N=B^2=I$, $ABAB=I$.\smallskip}

{\narrower\smallskip
\noindent{\it Prop.2$^\prime$: $S_N$ Invariance of (13).}

\noindent There is an $S_N$ action on solutions
of (13); the action of the fundamental transpositions $T_i=(i~i+1)$,
$i=1,..,N-1$, is given by
$$\eqalign{K_i&\rightarrow K_i+
   {{z(\alpha_{i+1}-\alpha_i-{\textstyle{1\over 2}})}
      \over{K_i+K_{i+1}+{\textstyle{1\over 2}}z}} \cr
           K_{i+1}&\rightarrow K_{i+1}-
   {{z(\alpha_{i+1}-\alpha_i-{\textstyle{1\over 2}})}
      \over{K_i+K_{i+1}+{\textstyle{1\over 2}}z}} \cr
           K_r&\rightarrow K_r,~~r\not=i,i+1 \cr
           \alpha_i&\rightarrow\alpha_{i+1}-{\textstyle{1\over 2}}   \cr
           \alpha_{i+1}&\rightarrow\alpha_{i}+{\textstyle{1\over 2}}    \cr
           \alpha_r&\rightarrow\alpha_r,~~r\not=i,i+1.  }\eqno{(15)}$$
\smallskip}

\noindent The transformations $T_i$ can easily be checked using the
reduced form of (5)$^\prime$ (equivalent to (13)):
$$ 2z(\dot{K}_i+\dot{K}_{i+1})=(K_{i+1}+K_i)(K_{i+1}-K_i+1)
       +{\textstyle{1\over 2}}z(K_{i+1}-K_i)+z(\alpha_i-\alpha_{i+1}),
    ~~~~~~~i=1,..,N-1.           \eqno{(13)'}$$
For comparison with [6],[9] it is useful to define
$\beta_i\equiv N^{-1}(2\alpha_i-i+{1\over 2}(N+1))$; the action of $T_i$,
$i=1,..,N-1$, on the $\beta_r$
is $\beta_i\rightarrow\beta_{i+1}$, $\beta_{i+1}\rightarrow\beta_i$,
and $\beta_r\rightarrow\beta_r$ for $r\not=i,i+1$. The action of $T_N$
is $\beta_1\rightarrow\beta_N+1$, $\beta_N\rightarrow\beta_1-1$,
and $\beta_r\rightarrow\beta_r$ for $r\not=1,N$. The action of $A$ is
$\beta_r\rightarrow\beta_{r+1}+{1\over N}$ for $r\not=N$, and $\beta_N
\rightarrow\beta_1-{{N-1}\over N}$, and the action of $B$ is
$\beta_r\rightarrow -\beta_{N+1-r}$. The transformation $AT_1T_2...T_{N-1}$
acts as a ``parallel transformation'', mapping $\vec{\beta}$ to
$\vec{\beta}+{1\over N}(1,1,..,1,1-N)$.

We now relate the above systems for $N=3,4$ to \P4 and \P5, and discuss
the relevant transformation groups. The following results are elementary
to establish.

{\narrower\smallskip
\noindent{\it Prop.3}

\noindent The general solution of (13) for $N=3$ is
$$\eqalign{K_3&={{M+z}\over 2}\cr
       K_2-K_1&={{2z\dot{M}}\over M}-1
                +{{z(1+2\alpha_1-2\alpha_2)} \over {M}}
          }\eqno{(16)}$$
where $M(z)$ solves the equation
$$ \ddot{M}={{\dot{M}^2}\over{2M}}+
    {{3M^3}\over{32z^2}}+{{3M^2}\over {8z}}
    +\left({3\over 4}-{{2\alpha_3}\over z}-{1\over{z^2}}\right){3M\over 8}
    -{{({\textstyle{1\over 2}}+\alpha_1-\alpha_2)^2}\over {2M}}.
   \eqno{(17)}$$
$M(z)$ solves (17) if and only if  $J(p)$ defined by $M(z)=2pJ(p)$, where
$p=(3z/4)^{1\over 2}$, satisfies \P4:
$$ {{d^2J}\over{dp^2}}={1\over {2J}}\left({{dJ}\over{dp}}\right)^2
     +{\textstyle{3\over 2}}J^3+4pJ^2+2(p^2-2\alpha_3)J
     -{2\over J}\left({{1+2\alpha_1-2\alpha_2}\over 3}\right)^2. \eqno{(18)}$$
\smallskip}

\noindent The transformation group for (17) is a semi-direct product
of the group generated by $T_1,T_2,T_3$, which Okamoto [6] calls $s_1,s_2,
\tilde{s}$, with the $D_N$ group generated by $A,B$. Okamoto writes
$\tilde{l}$ for $A^{-1}$ (and $l$ for $T_2T_1A^{-1}$),
and instead of $B$ uses $x=AB$ (all this can
easily be checked; Okamoto's coefficients $v_1,v_2,v_3$ are the
coefficients $\beta_1,\beta_2,\beta_3$ introduced above). The transformation
group for (18) is just that for (17) supplemented with the extra
symmetry $J\rightarrow -J$, $p\rightarrow -p$,  which Okamoto denotes $\psi$.
Explicit formulae for all the transformations can easily be written.

{\narrower\smallskip
\noindent{\it Prop.4}

\noindent The general solution of (13) for $N=4$ is
$$\eqalign{
K_1&= {{z\dot{V}}\over{V(V-1)}} + {z\over 4}{{V+1}\over{V-1}}
      -{{V+1+2(V-1)(\alpha_2-\alpha_1)}\over{4V}} \cr
K_2&= -{{z\dot{V}}\over{V(V-1)}} + {z\over 4}{{V+1}\over{V-1}}
      +{{V+1+2(V-1)(\alpha_2-\alpha_1)}\over{4V}} \cr
K_3&=  {{z\dot{V}}\over{(V-1)}} - {z\over 4}{{V+1}\over{V-1}}
      -{{V+1+2(V-1)(\alpha_3-\alpha_4)}\over{4}} \cr
}\eqno{(19)}$$
where $V(z)$ solves \P5:
$$\eqalign{
  \ddot{V}=\left({1\over{2V}}+{1\over{V-1}}\right)\dot{V}^2-{{\dot{V}}\over z}
    &-{{V(V+1)}\over{8(V-1)}}+{{(\alpha_1+\alpha_2)V}\over{2z}}\cr
    &+{{(V-1)^2}\over{32z^2}}\left((1+2\alpha_3-2\alpha_4)^2V-
                                  {{(1+2\alpha_1-2\alpha_2)^2}\over V}\right).
     }\eqno{(20)}$$
\smallskip}

\noindent Note that for $N\not=4$  the order of the system (13) is $N-1$,
but for $N=4$ it is 2. The form of \P5 in (20) is brought to the standard
form of Okamoto [9] by rescaling $z$. Having done this, it is straightforward
to check that the coefficients $\beta_1,..,\beta_4$ are exactly the
coefficients $v_1,..,v_4$ of Okamoto. The relationship between the
transformations we have introduced and those of [9] is as follows.
First we note that since $V$ is determined by $K_1+K_2$, the transformations
$T_1,T_3$ leave $V$ unchanged. These are $\pi_1^\prime$ and $\pi_1$ in [9],
respectively. $s_1,s_2,s_3,s_0$ in [9] are $T_1,T_2,T_3,T_4$ respectively,
and $l$ is $T_3T_2T_1A^{-1}$. Finally $x$ (or $\pi_2$) of [9] is $BA^2$,
and $w'$ of [9] is $T_1T_3B$.

\vskip.3in

\noindent{\bf Discussion}

It is pleasing that we have obtained the results of [6] and [9]
in a unified and extended framework; we see the rather
complicated transformation groups for \P4 and \P5 have a fairly
simple origin in the symmetries of (5).
It is to be expected that useful applications will be found
for the system (13) for $N\ge 5$, and for the systems obtained
by scaling reduction of higher equations in the MKdV hierarchies
(all of these systems possess the Painlev\'e property).
For example, in generalization of the results of [11], we should expect
the scaling reductions of equations in the $SL(N)$ MKdV hierarchy to
arise as the ``string equations'' for certain matrix models.

One thing missing from this paper is an explanation of the
origin of the transformation groups for \P3 [12] and \P6 [13]. From
Okamoto's work on these systems one might guess that they
arise as scaling reductions of an MKdV equation associated with
the Lie algebras $B_2$ and $D_4$ respectively [14].
The lowest nontrivial flow in the
$B_2$ MKdV hierarchy (which describes the evolution of two fields
$j_1,j_2$) can be computed, and has a consistent scaling
reduction; each of the resulting pair of equations can be integrated once,
as above, but the remaining system is a {\it fourth} order system with
two arbitrary constants. Remarkably, this system can actually be written
as a single fourth order ODE. But so far I have been unable to establish
any connection between this equation and \P3. From [5] it is clear
that \P3 and \P6 have to be discussed in tandem with other systems, so
it might not be surprising if they arose naturally embedded in some
larger system; but currently the existence of a relation between the
$B_2$ MKdV and \P3 remains conjecture.

Another possibility  for the origin of the transformation groups of \P3
and \P6 is that these equations might arise as scaling reductions of
some other bihamiltonian integrable system in $1+1$ dimensions (it is
known that \P3 and \P6  arise as reductions of
certain $1+1$ dimensional systems - see [15], p.343 for references -
but not as scaling reductions). It can be shown that group actions
which survive scaling reduction exist on the spaces
of solutions of other bihamiltonian systems. Indeed the reader can check
that \P4 arises as a scaling reduction of the system
$$ \eqalign{j_t&=(j_x+j^2-2j\bar{j})_x\cr
            \bar{j}_t&=(-\bar{j}_x-\bar{j}^2+2j\bar{j})_x}\eqno{(21)}$$
which is intimately related to the nonlinear Schr\"odinger equation
(see [16]). Simple
B\"acklund transformations of (21), such as
$j\rightarrow\bar{j},\bar{j}\rightarrow j,x\rightarrow x,
t\rightarrow -t$ generate at least some of the
transformations for \P4 (compare [17] where some of the transformations
for \P4 were obtained directly from B\"acklund transformations of NLS).

As a final comment, I note that a new derivation of all the
Painlev\'e equations has recently been given by Mason and Woodhouse,
who examined certain symmetry reductions of the $SL(2)$
self-dual Yang-Mills equations [18]. It would be very interesting to see
if the transformation groups had an explanation from this viewpoint as well.

\vskip.3in
\noindent{\bf Acknowledgements}

\smallskip
\noindent I thank P.Aspinwall and
C.Johnson for discussions. This work was supported by
the U.S.Department of Energy under grant \#DE-FG02-90ER40542.

\vskip.3in
\noindent{\bf References}

\smallskip
\noindent
\item{[1]} G.L.Lamb Jr.,{\it Elements of Soliton Theory}, Wiley (1980).
\item{[2]} J.Weiss, M.Tabor and G.Carnevale, {\it J.Math.Phys.} {\bf 24}
  (1983) 522.
\item{[3]} J.Weiss, in {\it Painlev\'e Transcendents, Their Asymptotics
and Physical Application}, eds. D.Levi and P.Winternitz, Plenum (1992).
\item{[4]} M.Boiti and F.Pempinelli, {\it Nuov.Cim.B} {\bf 51} (1979) 70.
\item{[5]} A.S.Fokas and M.J.Ablowitz, {\it J.Math.Phys.} {\bf 23} (1982) 2033.
\item{[6]} K.Okamoto, {\it Math.Ann.} {\bf 275} (1986) 221.
\item{[7]} G.R.W.Quispel, J.A.G.Roberts and F.W.Nijhoff, {\it Phys.Lett.A}
  {\bf 91} (1982) 143.
\item{[8]} Y.Murata, {\it Funk.Ekvac.} {\bf 28} (1985) 1.
\item{[9]} K.Okamoto, {\it Japan J.Math.} {\bf 13} (1987) 47.
\item{[10]} G.Wilson, {\it Nonlinearity} {\bf 5} (1992) 109.
\item{[11]} S.Dalley, C.V.Johnson, T.R.Morris and A.W\"atterstam,
       {\it Mod.Phys.Lett.A} {\bf 7} (1992) 2753.
\item{[12]} K.Okamoto, {\it Funk.Ekvac.} {\bf 30} (1987) 305.
\item{[13]} K.Okamoto, {\it Ann.Math.Pur.Appl.} {\bf 146} (1987) 337.
\item{[14]} V.G.Drinfeld and V.V.Sokolov, {\it Jour.Sov.Math.} {\bf 30}
     (1985) 1975.
\item{[15]} M.J.Ablowitz and P.A.Clarkson, {\it Solitons, Nonlinear
   Evolution Equations and Inverse Scattering}, Cambridge (1991).
\item{[16]} J.Schiff {\it The Nonlinear Schr\"odinger Equation and
    Conserved Quantities in the Deformed Parafermion and SL(2,R)/U(1)
    Coset Models}, IAS preprint, IASSNS-HEP-92/57 (Aug 1992).
\item{[17]} M.Boiti and F.Pempinelli, {\it Nuov.Cim.B} {\bf 59} (1980) 40.
\item{[18]} L.J.Mason and N.M.J.Woodhouse, {\it Self-duality and the
    Painlev\'e transcendents}, Oxford preprint (Nov 1992).

\vskip.3in
\noindent{\bf Appendix: Derivation of (5)}

\smallskip

The self-dual Yang-Mills equations can be written $F_{\bar{x}\bar{t}}=0$,
$F_{x\bar{t}}=F_{t\bar{x}}$, $F_{xt}=0$, where
$F_{\mu\nu}=\partial_\mu A_\nu-\partial_\nu A_\mu+[A_\mu,A_\nu]$ ($\mu,\nu\in
\{x,t,\bar{x},\bar{t}\}$) and the $A_\mu$ are the ``potentials''
i.e. Lie-algebra valued functions of $x,t,\bar{x},\bar{t}$.
Equations (5), up to a rescaling of $t$,
are obtained from the $SL(N)$ self-dual Yang-Mills
equations with an ansatz:
$$\eqalign{
(A_{\bar{x}})_{ij}&=\delta_{i,j+(N-1)}  \cr
(A_{\bar{t}})_{ij}&=f\delta_{i,j+(N-1)} - \delta_{i,j+(N-2)} \cr
(A_x)_{ij}        &=j_i\delta_{i,j} + \delta_{i,j-1} \cr
(A_t)_{ij}        &=A_i\delta_{i,j} + B_i\delta_{i,j-1}
                         - \delta_{i,j-2} \cr  }$$
Here the $A_i$ ($i=1,..,N$), $B_i$ ($i=1,..,N-1$), $j_i$ ($i=1,..,N$)
and $f$ are functions of $x,t$ alone, with $\sum_{i=1}^N j_i=0$
and $\sum_{i=1}^N A_i=0$.

\bye